\documentclass[nofootinbib,twocolumn,showpacs,preprintnumbers,pre,aps]{revtex4-1}

\usepackage{graphicx}
\usepackage{bm}
\usepackage{amsmath}
\usepackage{amssymb}
\usepackage{color}

\begin{document}

\title{Coexistences of lamellar phases in ternary surfactant solutions}%

\author{Isamu Sou}

\author{Ryuichi Okamoto}

\author{Shigeyuki Komura}\email{komura@tmu.ac.jp}

\affiliation{
Department of Chemistry, Graduate School of Science and Engineering,
Tokyo Metropolitan University, Tokyo 192-0397, Japan}

\author{Jean Wolff}

\affiliation{Institut Charles Sadron, UPR22-CNRS 23,
rue du Loess BP 84047, 67034 Strasbourg Cedex, France}

\date{\today}

\begin{abstract}
We theoretically investigate the coexistences of lamellar phases both in binary and ternary 
surfactant solutions.  
The previous free energy of a lamellar stack is extended to take into account the translational 
entropy of membrane segments. 
The obtained phase diagram for binary surfactant solutions (surfactant/water mixtures)
shows a phase separation between two lamellar phases and also exhibits a critical point. 
For lamellar phases in ternary surfactant solutions (surfactant/surfactant/water mixtures),
we explore possible phase behaviors and show that the phase diagrams exhibit various 
three-phase regions as well as two-phase regions in which different lamellar phases coexist.
We also find that finite surface tension suppresses undulation fluctuations of membranes 
and leads to a wider three-phase and two-phase coexistence regions.
\end{abstract}

\maketitle

\section{Introduction}
\label{sec:introduction}

One of the simplest mesoscale structures found in mixtures of water and surfactant molecules 
is the lamellar phase in which bilayers of amphiphilic molecules form roughly parallel sheets 
separated by water~\cite{SO}. 
In some surfactant/water binary systems, it is known that the lamellar phase can be swollen 
almost without limit. 
For example, in the mixture of  C$_{12}$E$_5$ ($n$-akylpolyglycolether) and water, 
the repeat distance of the lamellar phase can exceed 3,000~\AA~\cite{SSRNO}. 
The transition from the bound lamellar phase to the unbound phase is generally called the 
``unbinding transition"~\cite{LS}.
Although it is very rare, the coexistence of two lamellar phases in thermodynamic equilibrium 
has been also reported for binary surfactant/water solutions typically containing  
DDAB (didodecyldimethylammonium bromide)~\cite{Dubois92,Zemb93,Dubois98,Brotons05}.
In this case, a higher-density condensed lamellar phase is in equilibrium with a lower-density 
swollen lamellar phase. 
However, the reason why such a lamellar-lamellar coexistence is so rare in binary surfactant 
solutions is still a matter of debate and even puzzling~\cite{Silva07,Silva10}.

In contrast to binary mixtures, lamellar-lamellar coexistence is fairly common in ternary systems 
such as surfactant/surfactant/water
mixtures~\cite{Lis81a,Lis81b,Marques93,McGrath97,Ricoul98,Montalvo02} 
or polymer/surfactant/water mixtures~\cite{Rong96,Deme97,Bryskhe01}. 
Here the surfactant molecules assemble into stacked bilayers, while they are organized as 
coexisting lamellar phases.
The fact that ternary solutions typically exhibit a lamellar-lamellar coexistence can be explained 
if the bilayers with different components have different interactions across the water 
layer~\cite{Noro99}. 
In such a lamellar-lamellar phase separation, it is known that a long-ranged repulsive interaction 
such as electrostatic interaction and/or steric repulsive interaction play an important role. 
The latter interaction is known as the Helfrich steric interaction which arises 
from the reduced undulation entropy of fluctuating membranes~\cite{Helfrich78,SafranBook}.
In other words, the excluded volume of the neighboring membranes limits the configuration 
of a membrane and hence reduce its entropy.

For example, Harries \textit{et al.}\ investigated phase separations of charged surfactants by 
taking into account both the electrostatic and non-electrostatic interactions within a mean-field 
theory~\cite{Harries06}. 
They found that the lamellar-lamellar phase separation is controlled by non-electrostatic 
interactions between the counterions, and also by the interactions between the neutral and 
charged surfactants.
On the other hand, lamellar-lamellar coexistences in charged membranes were described only by 
electrostatic interactions in the other work~\cite{Jho10}.

For electrically neutral bilayer membranes, the combination of the steric repulsive interaction and 
other direct microscopic interactions, such as long-ranged van der Waals attraction and short-ranged 
hydration repulsion, determines whether membranes bind each other or unbind to have an 
infinite separation between them.
Lipowsky and Leibler pointed out that a simple superposition of the Helfrich steric 
repulsion and other direct interactions within a mean-field level gives incorrect (first-order)  description 
of the unbinding transition~\cite{LL1}.
An appropriate treatment of this problem using a functional renormalization-group method showed 
that the unbinding transition should be a continuous second-order transition
(known as the \textit{critical} unbinding transition).

\begin{figure}[tbh]
\begin{center}
\includegraphics[scale=0.3]{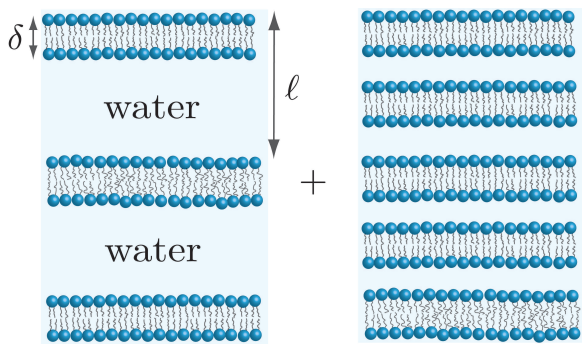}
\end{center}
\caption{
Schematic representation of two coexisting lamellar phases in binary 
surfactant/water solutions.
$\ell$ and $\delta$ are the lamellar repeat distance and the bilayer thickness, respectively.
For such a lamellar structure,  the surfactant volume fraction is given by $\phi=\delta/\ell$. 
The dilute lamellar phase (left) is characterized by a smaller $\phi$ value, while the condensed 
one (right) has a larger $\phi$ value.
}
\label{singlepicture}
\end{figure}

Later, Milner and Roux proposed a theory for the unbinding transition in a bulk of lamellar phase 
following the spirit of a mean-field theory for polymers~\cite{MR}.
In their argument, the Helfrich estimate of the entropy is taken into account accurately, whereas 
the other direct microscopic interactions are approximately incorporated as a correction to the 
hard-wall result for the second virial coefficient.
Their theory correctly accounts for the second-order nature of the critical unbinding transition.
Furthermore, it has been used to predict both the unbinding and preunbinding behaviors of a 
lamellar stack in binary surfactant solutions~\cite{Komura06}.

In this paper, we investigate the coexistences of lamellar phases both in binary surfactant 
solutions (surfactant/water mixtures) and ternary surfactant solutions 
(surfactant/surfactant/water mixtures) within a mean-field theory. 
We consider a situation when the surfactant molecules are electrically neutral, or the 
electrostatic interaction is sufficiently screened in the presence of electrolyte.
By taking into account the translational entropy of membrane segments,  we extend the 
mean-field theory by Milner and Roux in order to properly account for the phase behaviors in  
binary and ternary surfactant solutions.

Based on the proposed phenomenological free energy, we first discuss the phase diagrams of  
binary systems in which we find a lamellar-lamellar coexistence that ends in a critical point, 
as found in one of the experiments~\cite{Zemb93}.
By considering three interaction parameters (virial coefficients) between different components, 
we further discuss ternary mixtures and explore possible types of ternary phase diagrams 
(Gibbs triangles).
The phase behavior is very rich, and three-phase coexistences as well as two-phase coexistences
between different lamellar phases are predicted for a certain range of the interaction parameter 
values. 
We investigate both symmetric and asymmetric cases in terms of the two surfactant/water 
interactions.
In the symmetric case, the interactions (or the virial coefficients) between the 
same surfactant species are identical, whereas they are different in the asymmetric case.
We also discuss the effects of finite membrane surface tension on the phase behavior of 
ternary surfactant solutions.

\begin{figure}[tbh]
\begin{center}
\includegraphics[scale=0.3]{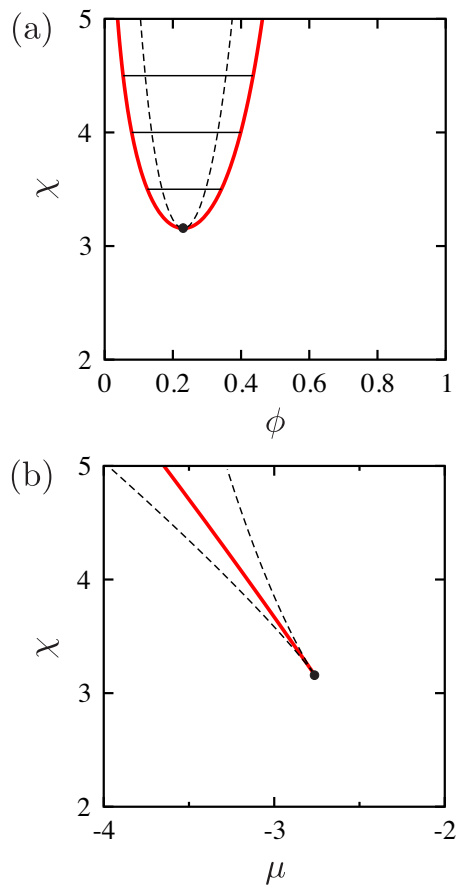}
\end{center}
\caption{
Phase diagrams of a binary surfactant solution as a function of (a) the surfactant volume 
fraction $\phi$ and the (scaled) virial coefficient $\chi$,  and (b) the surfactant chemical 
potential $\mu$ and the virial coefficient $\chi$. 
The red solid lines are binodal lines, and the black solid lines are tielines separating two distinct 
lamellar phases.
The black dashed lines are spinodal lines. 
The black circles represent the critical point at $(\chi_{\rm c}, \phi_{\rm c}, \mu_{\rm c})
=(3.16, 0.23, -2.76)$.
}
\label{single}
\end{figure}

In the next section, we explain the extension of the mean-field theory by Milner and Roux. 
Using the extended free energy, we first calculate the phase diagrams of binary surfactant 
solutions.
In Sec.~\ref{sec:two}, we consider the phase behavior of ternary surfactant solutions.
Various types of ternary phase diagrams are obtained both for the symmetric and asymmetric
cases.
In Sec.~\ref{sec:tension}, we also calculate the ternary phase diagrams in the presence of 
finite surface tension acting on the membranes. 
The summary of our work and some discussions are given in the last Sec.~\ref{sec:discussion}.

\section{Lamellar phases in binary mixtures}
\label{sec:single}

\begin{figure*}[tbh]
\begin{center}
\includegraphics[scale=0.4]{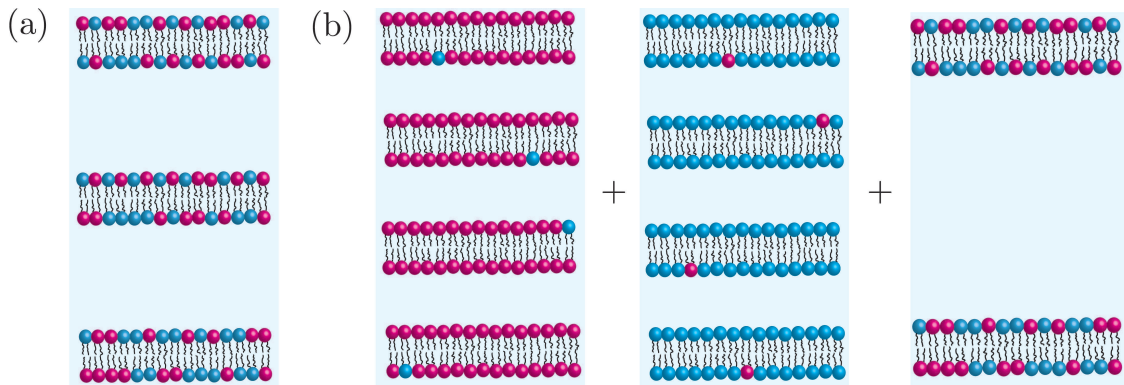}
\end{center}
\caption{
Schematic representation of lamellar phases in ternary surfactant solutions, i.e.,
surfactant/surfactant/water mixtures.
Each bilayer consists of surfactant A (blue) and surfactant B (red) whose volume 
fractions are $\phi$ and $\psi$, respectively. 
(a) A single lamellar phase having a unique repeat distance that is fixed by the total 
surfactant volume fraction $\phi+\psi$.
(b) An example of three coexisting lamellar phases characterized either by different 
repeat distances or A/B relative compositions. 
}
\label{threephase}
\end{figure*}

Bilayer fluid membranes experience steric repulsion arising from their reduced undulation 
entropy~\cite{SO}.
The corresponding interaction energy \textit{per unit area} of membrane was considered 
by Helfrich and is given by~\cite{Helfrich78,SafranBook}
\begin{equation}
v_{\rm s}(\ell) = \frac{b (k_{\rm B}T)^2}
{\kappa (\ell-\delta)^2}.
\label{steric}
\end{equation}
Here $k_{\rm B}$ is the Boltzmann constant, $T$ is the temperature,
$\kappa$ is the bending rigidity, $\ell$ is the average repeat distance between 
bilayers, as shown in Fig.~\ref{singlepicture}, and a constant $\delta$ is the 
membrane thickness that is used as the smallest cutoff length. 
Note that $\ell-\delta$ in the denominator corresponds to the inter-membrane distance 
in which membranes can undergo out-of-plane fluctuations.
The numerical prefactor $b$ was calculated to be $b=3\pi^2/128 \approx 0.23$ 
in the original work by Helfrich~\cite{Helfrich78}, but its value is debatable in the 
literatures~\cite{LZ,NL93,NL95}.
For example, Monte Carlo simulations in Ref.~\cite{NL95} yielded a lower value of $b\approx 0.12$, 
almost a half of the above value.
In the present study, the exact value of $b$ does not affect the results because we rescale 
all the energy densities by including the factor $b$ [see later Eq.~(\ref{scaleMR})].

In order to describe the free energy of a lamellar stack in a binary surfactant/water solution, 
we first introduce the membrane volume fraction $\phi = \delta/\ell \ge 0$.
Here we have assumed that all the surfactant molecules constitute bilayers.
Extending the argument by Milner and Roux~\cite{MR}, we consider the following grand potential 
\textit{per unit volume} of a lamellar stack: 
\begin{align}
\bar{f}(\phi) & = \frac{k_{\rm B} T}{\delta^3} \phi (\log \phi -1) 
- k_{\rm B}T \bar{\chi} \phi^2 
\nonumber \\
& + \frac{b (k_{\rm B}T)^2}{\kappa \delta^3} 
\frac{\phi^3}{(1-\phi)^2} - \bar{\mu} \phi. 
\label{milnerroux}
\end{align}
Here the first term represents the translational entropy of membrane segments, 
which was not considered before~\cite{MR}.  
This term, however, plays an essential role when we calculate the phase 
diagrams within equilibrium thermodynamics. 
In general, the translational entropy term should be given by  
$\phi \log \phi + (1-\phi) \log (1-\phi)$~\cite{SafranBook,DoiBook}, 
and the first term in Eq.~(\ref{milnerroux}) corresponds to its lowest order 
expansion in terms of small $\phi \ll 1$. 
However, since such a generalization does not result in any essential modification, 
we shall study the above grand potential in this paper.
Our approximation is justified because the term $(1-\phi) \log (1-\phi)$ vanishes
when $\phi \rightarrow 1$, while the third term in Eq.~(\ref{milnerroux}) diverges, 
as explained below.

The second term in Eq.~(\ref{milnerroux}) is the correction to the entropic hard-wall result, 
and $\bar{\chi}$ is the second virial coefficient obtained from  
\begin{equation}
\bar{\chi} = - \frac{1}{2\nu^2} \int {\rm d} \mathbf{r}\, 
(1 - \exp[-U_{\nu}(\mathbf{r})/k_{\rm B}T]),
\label{virial}
\end{equation}
where $\nu \approx \delta^3$ is the volume of the membrane segment, and $U_{\nu}(\mathbf{r})$ 
is the interaction between bits of membrane of volume $\nu$, and $\mathbf{r}$ is a three-dimensional vector.
All the direct microscopic (van der Waals, hydration, electrostatic) interactions are taken into 
account through $U_{\nu}(\mathbf{r})$.
The third term in Eq.~(\ref{milnerroux}) is due to the Helfrich steric repulsion, and the factor of 
$(1-\phi)^{-2}$ comes from the finite membrane thickness [see also Eq.~(\ref{steric})]~\cite{DC}.
Finally,  the chemical potential, $\bar{\mu}$, is needed for the conservation of 
the surfactant volume fraction $\phi$.
A similar free energy to Eq.~(\ref{milnerroux}) was also proposed in other 
works to describe the unbinding transition~\cite{Helfrich89,Komura03,Bougis15} but without 
the translational entropy term that we have introduced.
We note again that the first translational entropy term in Eq.~(\ref{milnerroux}) is not accounted 
for by the Helfrich steric repulsion term which also has an entropic origin. 
Without the translational entropy term, the behavior of the free energy is thermodynamically 
inappropriate around $\phi \approx 0$ and one cannot describe correct phase behaviors.

\begin{figure}[tbh]
\begin{center}
\includegraphics[scale=0.4]{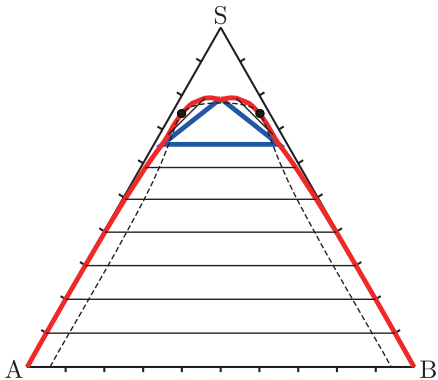}
\end{center}
\caption{
Phase diagram of a ternary surfactant solution when the interaction
parameters are  $\chi_{\phi\phi}=\chi_{\psi\psi}=3$ and $\chi_{\phi\psi}=-3$.
``A" and ``B" stand for surfactant A and surfactant B, respectively, while ``S" indicates 
solvent such as water.
The phase diagram is symmetric with respect to the equal A/B relative compositions.
The red solid lines are binodal lines, and the black solid lines are tielines separating two 
distinct lamellar phases.
The blue triangle represents the region of three-phase coexistence, and the black dashed 
line is a spinodal line.
The upper white region corresponds to the one-phase lamellar region. 
The black circles represent the critical points at $(\phi_{\rm c}, \psi_{\rm c})=
(0.026, 0.23)$ and $(0.23, 0.026)$.
}
\label{chi=3}
\end{figure}

It is convenient to rescale all the energy densities by 
$2 b (k_{\rm B}T)^2/(\kappa \delta^3)$. 
Then Eq.~(\ref{milnerroux}) can be presented in a dimensionless form as 
\begin{equation}
f(\phi) =a \phi(\log \phi-1) - \chi \phi^2 + 
\frac{\phi^3}{2(1-\phi)^2}- \mu \phi,
\label{scaleMR}
\end{equation}
where $a=\kappa/(2bk_{\rm B}T)$ is a numerical factor of order unity, and hence 
can be set as $a=1$ in the following discussion for simplicity, while 
$\chi=\bar{\chi} \kappa \delta^3/(2bk_{\rm B}T)$ is a dimensionless interaction parameter.
The equation of state is then given by minimizing the grand potential, $\partial f/\partial \phi=0$, 
and becomes 
\begin{equation}
\mu =  \log \phi  -2 \chi \phi + \frac{3\phi^2}{2(1-\phi)^2}+\frac{\phi^3}{(1-\phi)^3}.
\label{state}
\end{equation}

Using the above grand potential density, we can obtain the spinodal from the condition~\cite{Koningsveld}: 
\begin{equation}
\frac{\partial^2 f(\phi)}{\partial \phi^2}=0.
\label{spinodal}
\end{equation}
With the use of Eq.~(\ref{scaleMR}), it can be written as 
\begin{equation}
\chi=\frac{1}{2}\left[\frac{1}{\phi}+\frac{3\phi}{(1-\phi)^2}+\frac{6\phi^2}{(1-\phi)^3}+\frac{3\phi^3}{(1-\phi)^4}\right].
\end{equation}
The conditions for the critical point is given by 
\begin{equation}
\frac{\partial^2 f(\phi)}{\partial \phi^2}=0,~~~~~~
\frac{\partial^3 f(\phi)}{\partial \phi^3}=0.
\label{binarycritical}
\end{equation}
These conditions and Eq.~(\ref{state}) can be numerically solved to obtain the 
critical point as $(\chi_{\rm c}, \phi_{\rm c},\mu_{\rm c})=(3.16, 0.23,-2.76)$.

On the other hand, the thermodynamic equilibrium between the two coexisting phases 
denoted as ``1'' and ``2'' and characterized by $\phi_1$ and $\phi_2$, satisfies the 
following conditions~\cite{Koningsveld}:
\begin{equation}
\frac{\partial f(\phi)}{\partial \phi} \biggr|_1 = \frac{\partial f(\phi)}{\partial \phi} \biggr|_2=0,~~~~~
f(\phi_1)=f(\phi_2). 
\label{binodal}
\end{equation}
We have numerically solved the above set of conditions to obtain the phase diagrams.

The calculated phase diagrams in the $(\chi,\phi)$ and $(\chi,\mu)$ planes are 
shown in Fig.~\ref{single}.
The red solid lines are binodal lines, and the black dashed lines are spinodal lines.
When $\chi > 3.16$, the binary mixture separates into two lamellar phases 
characterized by different $\phi$ values indicated by the horizontal tielines
(see also Fig.~\ref{singlepicture}). 
Notice that there is a critical point at $(\chi_{\rm c}, \phi_{\rm c}, \mu_{\rm c})=
(3.16, 0.23, -2.76)$ where the two lamellar phases become identical.
For $\chi < 3.16$, on the other hand, the \textit{complete} unbinding of the lamellae 
occurs upon swelling with excess water~\cite{LL2-1,LL2-2}.
In the $(\chi,\phi)$ phase diagram, such a transition occurs when we take the limit of 
$\phi \rightarrow 0$.

The calculated phase diagram in Fig.~\ref{single}(a) resembles that obtained for 
DDAB/water binary mixtures~\cite{Dubois92,Zemb93,Dubois98,Brotons05} if
we assume that the interaction parameter $\chi$ is inversely proportional to the
temperature.
In these experiments, the existence of a critical point and associated critical 
phenomena were experimentally evidenced by various scattering methods~\cite{Zemb93}.

\begin{figure}[tbh]
\begin{center}
\includegraphics[scale=0.4]{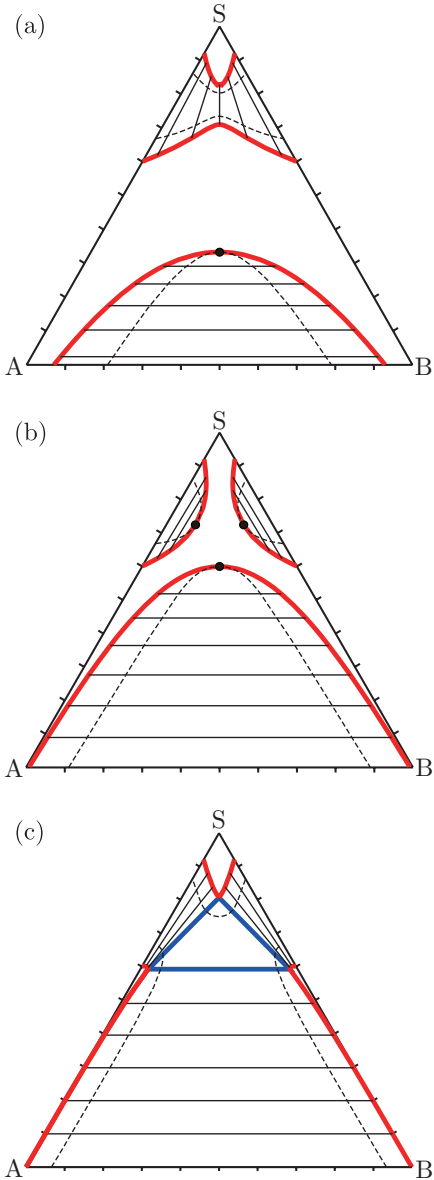}
\end{center}
\caption{
Phase diagrams of a ternary surfactant solution when the interaction
parameters are $\chi_{\phi\phi}=\chi_{\psi\psi}=4$ (symmetric) and 
(a) $\chi_{\phi\psi}=5$, 
(b) $\chi_{\phi\psi}=3$, 
(c) $\chi_{\phi\psi}=0$.
The meanings of different lines are explained in Fig.~\ref{chi=3}.
One and three critical points exist in (a) and (b) (black circles), respectively, while 
a three-phase coexistence region appears in (c) (blue triangle). 
}
\label{chi=4}
\end{figure}

\section{Lamellar phases in ternary mixtures}
\label{sec:two}

Next we consider lamellar phases in ternary surfactant/surfactant/water solutions in which 
bilayer membranes are composed of two different types of surfactant, say surfactant A and 
surfactant B, as shown in Fig.~\ref{threephase}.
Let us define the volume fractions of surfactant A and B by $\phi$ and $\psi$, respectively. 
Then the volume fraction of water (solvent) is automatically fixed by $1-\phi-\psi$ due to the 
incompressibility condition.
As a generalization of Eq.~(\ref{scaleMR}),  we consider the following dimensionless 
grand potential \textit{per unit volume} for ternary mixtures:
\begin{align}
g(\phi,\psi) & =  \phi(\log \phi-1) + \psi(\log \psi-1) 
\nonumber \\ 
&- \chi_{\phi\phi} \phi^2 - \chi_{\psi\psi}  \psi^2 -  \chi_{\phi\psi} \phi \psi 
\nonumber \\ 
& + \frac{(\phi+\psi)^3}{2(1-\phi - \psi)^2}- \mu_{\phi} \phi - \mu_{\psi} \psi.
\label{two-freeenergy}
\end{align}
In the above, the first two terms represent the translational entropy of each surfactant
component, the next three terms describe the different interactions characterized 
by the three dimensionless virial coefficients 
$\chi_{\phi\phi}$, $\chi_{\psi\psi}$, and $\chi_{\phi\psi}$ which are assumed  
as independent parameters.
The first term in the third line corresponds to the Helfrich steric repulsion acting between 
mixed membranes. 
Notice here that the total surfactant volume fraction $\phi+\psi$ corresponds to the volume 
fraction of membranes. 
Several other assumptions that lead to this expression are separately discussed in Sec.~\ref{sec:discussion}.
Furthermore, $\mu_{\phi}$ and $\mu_{\psi}$ are the chemical potentials for the two surfactants.
When, for example, surfactant B is absent and hence $\psi=0$, Eq.~(\ref{two-freeenergy}) reduces to 
Eq.~(\ref{scaleMR}), as it should.

\begin{figure}[tbh]
\begin{center}
\includegraphics[scale=0.4]{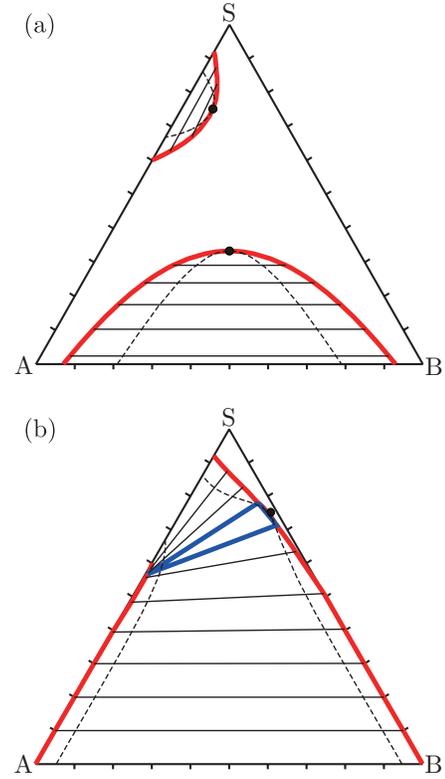}
\end{center}
\caption{
Phase diagrams of a ternary surfactant solution when the interaction
parameters are
(a) $\chi_{\phi\phi}=4$, $\chi_{\psi\psi}=3$, $\chi_{\phi\psi}=4$ and 
(b) $\chi_{\phi\phi}=4$, $\chi_{\psi\psi}=3$, $\chi_{\phi\psi}=-3$.
These phase diagrams are asymmetric with respect to the equal A/B relative compositions.
The meanings of different lines are explained in Fig.~\ref{chi=3}.
Two and one critical points exist in (a) and (b) (black circles), respectively, while 
a three-phase coexistence region appears in (b) (blue triangle). 
}
\label{chi=4&3}
\end{figure}

In order to discuss the stability of the above grand potential with two independent variables, 
we consider the $2\times 2$ Hessian matrix of $g(\phi,\psi)$ given by~\cite{Koningsveld}
\begin{eqnarray}
\mathbf{H} =\left(
\begin{array}{cc}
\dfrac{\partial^2 g}{\partial \phi^2} & \dfrac{\partial^2 g}{\partial \phi \partial \psi} \\[2ex]
\dfrac{\partial^2 g}{\partial \psi \partial \phi} & \dfrac{\partial^2 g}{\partial \psi^2} 
\end{array}
\right).
\label{eq:bb}
\end{eqnarray}
At the spinodal, the Hessian defined as the determinant of the matrix, $H=\det \mathbf{H}$, 
vanishes, i.e., $H=0$.  
The critical point can be obtained by considering another $2\times2$ matrix
\begin{eqnarray}
\mathbf{H}' &=&\left(
\begin{array}{cc}
\dfrac{\partial^2 g}{\partial \phi^2} & \dfrac{\partial^2 g}{\partial \phi \partial \psi} \\[2ex]
\dfrac{\partial H}{\partial \phi} & \dfrac{\partial H}{\partial \psi}
\end{array}
\right),
\end{eqnarray}
and its determinant $H'=\det \mathbf{H}'$. 
Then the conditions for the critical point are given by~\cite{Koningsveld}
\begin{equation}
H=0,~~~~~H'=0.
\label{cri}
\end{equation}

For ternary mixtures, the thermodynamic equilibrium between the two coexisting phases denoted 
as ``1" and ``2" and characterized by $(\phi_1, \psi_1)$ and $(\phi_2, \psi_2)$, satisfies the conditions~\cite{Koningsveld}:
\begin{equation}
\frac{\partial g(\phi, \psi)}{\partial \phi} \biggr|_1 = 
\frac{\partial g(\phi, \psi)}{\partial \phi} \biggr|_2=0, 
\end{equation}
\begin{equation}
\frac{\partial g(\phi, \psi)}{\partial \psi} \biggr|_1 = 
\frac{\partial g(\phi, \psi)}{\partial \psi} \biggr|_2=0, 
\end{equation}
\begin{equation}
g(\phi_1, \psi_1)= g(\phi_2, \psi_2).
\end{equation}
Similarly, for a three-phase coexistence between phases ``1", ``2" and ``3", the following set 
of conditions should be satisfied~\cite{Koningsveld}:
\begin{equation}
\frac{\partial g(\phi, \psi)}{\partial \phi} \biggr|_1 = 
\frac{\partial g(\phi, \psi)}{\partial \phi} \biggr|_2= 
\frac{\partial g(\phi, \psi)}{\partial \phi} \biggr|_3=0,
\end{equation}
\begin{equation}
\frac{\partial g(\phi, \psi)}{\partial \psi} \biggr|_1 = 
\frac{\partial g(\phi, \psi)}{\partial \psi} \biggr|_2 =
\frac{\partial g(\phi, \psi)}{\partial \psi} \biggr|_3=0,
\end{equation}
\begin{equation}
g(\phi_1, \psi_1)= g(\phi_2, \psi_2)=g(\phi_3, \psi_3).
\end{equation}
For lamellar phases under consideration, an example of three-phase coexistence is schematically 
presented in Fig.~\ref{threephase}(b).
In the following, we present the numerically calculated ternary phase diagrams (Gibbs triangles)
for different interaction parameters.

We first consider the symmetric case between the two surfactants A and B, i.e.,
$\chi_{\phi\phi}=\chi_{\psi\psi}$.
Figure~\ref{chi=3} shows a ternary phase diagram when 
$\chi_{\phi\phi}=\chi_{\psi\psi}=3$ (symmetric) and $\chi_{\phi\psi}=-3$.
This is the case when each surfactant/water binary solution does not exhibit lamellar-lamellar 
phase separation because $\chi_{\phi\phi}= \chi_{\psi\psi}< 3.16$ 
(see Fig.~\ref{single}).  
The obtained phase diagram is always symmetric with respect the line $\phi=\psi$. 
Due to the strong repulsion between the A and B components ($\chi_{\phi\psi}=-3$), 
there is a wide region of two-phase coexistence (red solid lines) with horizontal tielines 
(black solid lines). 
In the upper part of the triangle, there is a region of three-phase coexistence (blue 
triangle) associated with two wings of two-phase coexistence. 
These two-phase coexistence regions end in two corresponding critical points (black circles).
The black dashes lines indicate the spinodal lines that appear inside the coexistence regions.
When we make $\chi_{\phi\psi}$ larger such as $\chi_{\phi\psi}=3$ (not shown), only the  
two-phase coexistence region remains in the lower part of the triangle with horizontal tielines, 
and the three-phase region disappears.

In Fig.~\ref{chi=4}, we present the ternary phase diagrams when $\chi_{\phi\phi}=\chi_{\psi\psi}=4$ 
(symmetric), while the A/B interaction is changed as (a) $\chi_{\phi\psi}=5$, (b) $\chi_{\phi\psi}=3$, 
and (c) $\chi_{\phi\psi}=0$.
These are the cases when each surfactant/water binary solution exhibits lamellar-lamellar 
phase separation because $\chi_{\phi\phi}= \chi_{\psi\psi}> 3.16$ (see Fig.~\ref{single}), 
as represent on the two S-A and S-B sides of the triangles. 
In the case of Fig.~\ref{chi=4}(a), these two-phase regions merge to form a single two-phase region
with tilted tielines in the upper part of the triangle. 
In the lower part of the triangle, on the other hand, there is a region of two-phase coexistence
with horizontal tielines. 
This two-phase region ends in a critical point. 
When $\chi_{\phi\psi}$ is made smaller as in Fig.~\ref{chi=4}(b), the upper two-phase coexistence region separates 
into distinct two-phase regions that also end in two corresponding critical points.  
For even smaller $\chi_{\phi\psi}$ as in Fig.~\ref{chi=4}(c), the three two-phase regions meet 
each other forming a three-phase coexistence region similar to Fig.~\ref{chi=3}. 
As a result, all the three critical points disappear.

In Fig.~\ref{chi=4&3}, we show the ternary phase diagrams for an asymmetric case 
of $\chi_{\phi\phi}=4>3.16$ and $\chi_{\psi\psi}=3<3.16$, while the A/B interaction is chosen as 
(a) $\chi_{\phi\psi}=4$ and (b) $\chi_{\phi\psi}=-3$.
In this case, only the binary A/S mixture exhibits the phase separation while the binary B/S 
does not. 
Here Fig.~\ref{chi=4&3}(a) should be compared with Fig.~\ref{chi=4}(b).
The two two-phase regions end in the receptive critical points.
When $\chi_{\phi\psi}$ is made smaller as in the case of Fig.~\ref{chi=4&3}(b), the two 
two-phase coexistence regions are connected to each other with the appearance of a 
three-phase coexistence region. 
This three-phase region accompanies another small two-phase region and a critical point
on the S-B side of the triangle.
In this asymmetric case, the tielines are not horizontal and tilted especially in the upper
part of the phase diagram. 　

\section{Effects of surface tension}
\label{sec:tension}

In this section, we consider lamellar phases in ternary surfactant solutions in which surface 
tension, $\sigma$, is acting on membranes.
It is known that finite surface tension significantly suppresses membrane undulations, and 
the range of fluctuation-induced interaction between tense membranes becomes shorter.
Although the calculation of this interaction is complicated in general, Seifert provided a 
simple self-consistent calculation~\cite{Seifert95}.
He showed that the energy \textit{per unit area} of membrane in the presence of 
surface tension is given by
\begin{equation}
v_{\rm s}(\ell;\xi) = \frac{b (k_{\rm B}T)^2}
{\kappa (\ell-\delta)^2}
\left[ 
\frac{(\ell - \delta)/\xi}{\sinh [(\ell - \delta)/\xi]}
\right]^2,
\label{tension-steric}
\end{equation}
where $\xi = (2 k_{\rm B}T/\pi \sigma)^{1/2}$ is the characteristic length arising from the 
competition between the thermal energy and the surface energy.
When the surface tension $\sigma$ is small ($\xi \rightarrow \infty$), the above expression reduces 
to Eq.~(\ref{steric}) for a tensionless membrane, and recovers the long-range algebraic decay.  
When the surface tension $\sigma$ is large ($\xi \rightarrow 0$), on the other hand, Eq.~(\ref{tension-steric}) 
decays exponentially with distance $\ell$, consistent with the renormalization-group result~\cite{LS} 
and Monte Carlo simulations~\cite{NL95}.
The reduction of membrane undulations in the presence of surface tension was experimentally 
observed in Ref.~\cite{Mutz89}.

\begin{figure}[tbh]
\begin{center}
\includegraphics[scale=0.4]{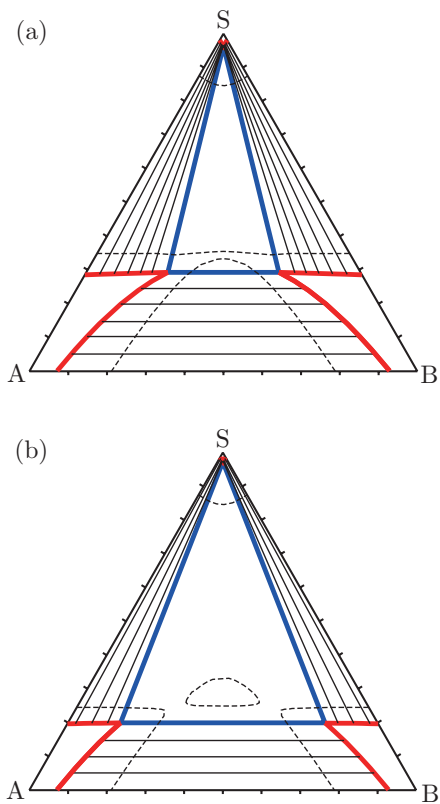}
\end{center}
\caption{
Phase diagram of a ternary surfactant solution in the presence of membrane surface tension. 
The interaction parameters are 
$\chi_{\phi\phi}=\chi_{\psi\psi}=4$ (symmetric), $\chi_{\phi\psi}=5$, 
while the tension parameters are 
(a) $\delta/\xi=10$ and (b) $\delta/\xi=20$. 
These phase diagrams should be compared with that in Fig.~\ref{chi=4}(a) for which  
$\delta/\xi=0$.
The wide three-phase coexistence region (blue triangle) further expands when $\delta/\xi$ 
becomes larger. 
}
\label{tensionx=3}
\end{figure}

Using Eq.~(\ref{tension-steric}) in the presence of surface tension, we consider a modified 
grand potential \textit{per unit volume} for ternary surfactant solutions as~\cite{DC} 
\begin{align}
g(\phi,\psi; x) & =  \phi(\log \phi-1) + \psi(\log \psi-1) 
\nonumber \\ 
&- \chi_{\phi\phi} \phi^2 - \chi_{\psi\psi}  \psi^2 -  \chi_{\phi\psi} \phi \psi 
\nonumber \\ 
& - x^2 G(x) \frac{(\phi+\psi)^3}{(1-\phi - \psi)^2}- \mu_{\phi} \phi - \mu_{\psi} \psi.
\label{tension-freeenergy}
\end{align}
Here the dimensionless quantity $x$ is defined by 
\begin{equation}
x=\left(\frac{\delta}{\xi} \right) \frac{1-\phi-\psi}{\phi+\psi},
\end{equation}
whereas the scaling function $G(x)$ is given by
\begin{equation}
G(x) = - \frac{1}{2 \sinh^2 x}. 
\end{equation}
We note that $x$ depends also on $\phi$ and $\psi$, while $\delta/\xi$ is a dimensionless
parameter that can be given externally.

Figure~\ref{tensionx=3} shows the calculated ternary phase diagrams when 
$\chi_{\phi\phi}=\chi_{\psi\psi}=4$ (symmetric) and $\chi_{\phi\psi}=5$, while 
the parameter controlling the surface tension is (a) $\delta/\xi=10$ and (b) $\delta/\xi=20$.
These phase diagrams should be compared with that in Fig.~\ref{chi=4}(a) for 
which $\delta/\xi=0$. 
As shown here with finite surface tension, the phase separation is dramatically enhanced 
and the upper two two-phase coexistence regions extend down to the middle part of the triangle.
At the expense of the critical point in Fig.~\ref{chi=4}(a), there appears a large three-phase 
coexistence region in the middle part.
This three-phase coexistence region is connected to the lower two-phase region with 
horizontal tielines. 
As we see in Fig.~\ref{tensionx=3}(b), the three-phase coexisting region further expands 
when $\delta/\xi$ is made larger. 
Therefore, surface tension promotes the phase separation between different lamellar phases.
The fact that dense lamellar phases coexist with an excess water on the two S-A and S-B sides 
of the triangles is in accordance with the experiment~\cite{Mutz89}.
In the presence of finite surface tension, they observed membranes merging one by one into 
bundles at mutually adhering membranes.
Also the water between the membranes was driven into a small number of compact water 
pockets~\cite{Mutz89}.

\section{Summary and discussion}
\label{sec:discussion}

In this paper, we have investigated the coexistences of lamellar phases both in binary 
and ternary surfactant solutions.  
To calculate the phase diagrams, we have extended the previous free energy of a 
lamellar stack~\cite{MR} by taking into account the translational entropy of membrane 
segments. 
The obtained phase diagrams for a binary surfactant solution show a phase separation 
between two lamellar phases and also exhibit a critical point. 
For lamellar phases in ternary surfactant solutions, we have further extended the free 
energy to take into account the different interactions between three species and explored 
possible phase behaviors.
The calculated phase diagrams include various coexistences between three different 
lamellar phases (three-phase regions) or between two lamellar phases (two-phase regions). 
A systematic change of the phase behavior has been observed by changing the 
interaction parameter between the two surfactant species.
Finally, we have looked at the effects of finite surface tension which suppresses  
membrane fluctuations and leads to a wider three-phase coexistence region.

We stress again that the addition of translational entropy terms in the free energies
[see Eqs.~(\ref{milnerroux}) and (\ref{two-freeenergy})] is essential in calculating 
the correct phase diagrams.
Without these terms, one cannot obtain the coexistence between two lamellar phases
having different repeat distances.
The original free energy by Milner and Roux was considered in order to explain the 
unbinding transition in surfactant solutions~\cite{MR}. 
Although their phase diagram exhibits a coexistence between a lamellar phase 
and excess water (i.e., unbound lamellar phase), a coexistence between two 
distinct lamellar phases does not occur.
We consider that these translational entropy terms should be included in addition 
to the Helfrich steric interaction which also has an entropic origin.

In the present work, we have assumed that the lamellar phase is the only lyotropic 
liquid crystaline phase that is formed for any temperature (interaction) and composition. 
In real surfactant solutions, however, this is certainly not the case, because typical 
phase diagrams contain other phases such as the micellar phase, the hexagonal 
phase, and the cubic phase~\cite{SSRNO}.
Rather than reproducing realistic phase diagrams by considering all the possible phases
in surfactant solutions, our purpose is to investigate in detail the competition between 
the Helfrich steric repulsion and other direct microscopic (van der Waals, hydration) 
interactions especially in ternary mixtures.
This is why we have extended the free energy of a lamellar stack by Milner and Roux,  
and calculated various coexistences only between the lamellar phases. 
A similar theoretical approach was made by Noro and Gelbart who also
discussed lamellar-lamellar phase separations in surfactant solutions~\cite{Noro99}.
We also note that the steric repulsive interaction acting between neighboring cylinders 
in the hexagonal phase is discussed in  Ref.~\cite{SO}, which can be used to extend 
our treatment.

Another simplification in our work is that we have not taken into account the composition 
dependence of the bending rigidity $\kappa$ or the surface tension $\sigma$ when 
a membrane is composed of two surfactants A and B. 
Usually, a membrane made of an A/B surfactant mixture will show an intermediate bending 
rigidity between those of pure membranes. 
In principle, such a change in the  bending rigidity affects the Helfrich steric interaction.
One of the possible ways to describe the intermediate behavior is to linearly interpolate 
between the two pure limits.
This approximation can be justified when the bending rigidities of the pure components
are not so different. 
When they are very different, such as in membranes composed of surfactant and amphiphilic 
polymer, a nonlinear effect on the bending rigidity becomes important~\cite{KS01}.
Furthermore, a detailed discussion on the surface tension in a mixed membrane was recently 
given by some of the present authors~\cite{Okamoto16}.

Although the phase diagrams calculated in this paper may not be simply compared with
experimentally obtained ones because of the previously mentioned reason, it is still useful 
to discuss the ternary phase diagrams of glycolipid/cationic surfactant/water mixtures 
at room temperature~\cite{Ricoul98}.
In this ternary mixture, they found the two-phase regions between the coexisting lamellar 
phases on both sides of the Gibbs triangle.  
More interestingly, they further identified two corresponding critical points and also the 
region of the three-phase coexistence according to the phase rule.
Such a phase behavior is very reminiscent to the phase diagrams in Fig.~\ref{chi=4}(b) and (c).
Although not yet done, we expect that one can reproduce the experimentally obtained 
phase diagram by further tuning the three interaction parameters in our model.

In this paper, we have considered a situation in which the surfactants are electrically 
neutral or the electrostatic interaction is sufficiently screened.
As is clear from Eq.~(\ref{steric}), the Helfrich steric repulsion is important only when 
the membrane is flexible, $\kappa \simeq k_{\rm B}T$. 
For strongly charged and unscreened membranes (no electrolyte), on the other hand, 
the dominant repulsion originates from the electrostatic interactions between flat 
membranes. 
The interplay between the electrostatics and fluctuations of a stack of membranes 
(without van der Waals and hydration forces) was studied before by Pincus 
\textit{et al.}~\cite{Pincus90}.  
When electrostatic interactions are strong enough (Gouy--Chapman regime) compared 
with the Helfrich steric repulsion, they showed that out-of-plane membrane fluctuations 
become smaller than the inter-membrane separation $\ell -\delta$.
In the other weaker electrostatic regimes, on the other hand, the suppression of membrane 
fluctuations is less important, and in some cases, the screened electrostatic interactions
can be completely neglected~\cite{Pincus90}.
Our assumption for charged membranes is justified for such situations.

Our results indicate that the lateral phase separation in mixed membranes causes 
different inter-membrane distances. 
Recently much efforts have been made to study the statics and dynamics of 
multi-component lipid membranes~\cite{Komura14}, mainly using giant unilamellar 
vesicles (GUVs) in the experiments~\cite{Smith09}.
In the future, it is interesting to study the phase behaviors of multi-lamellar vesicles 
composed of more than two types of lipid by taking into account the interactions 
between neighboring membranes. 
The present work would provide us with a useful theoretical guide for such a research direction.

\begin{acknowledgments}

We thank T.\ Kato and Y. Kawabata for useful discussions.
S.K. and R.O. acknowledge support from the Grant-in-Aid for Scientific Research on
Innovative Areas ``\textit{Fluctuation and Structure}" (Grant No.\ 25103010) from the Ministry
of Education, Culture, Sports, Science, and Technology of Japan, and the 
Grant-in-Aid for Scientific Research (C) (Grant No.\ 15K05250)
from the Japan Society for the Promotion of Science (JSPS).
J.W.\ acknowledges support from the C.N.R.S.\ (Centre National de la Recherche Scientifique), 
the French ORT association, and ORT school of Strasbourg.
\end{acknowledgments}


\end{document}